# Generalized Perfect Optical Vortex along Arbitrary Trajectories

Yue Chen[1], Tingchang Wang[2], Yuxuan Ren[3], Zhaoxiang Fang[4], Guangrui Ding[2], Liqun He[5], Rongde Lu[6,7], Kun Huang[4,8]

[1]Department of Engineering and Applied Physics, University of Science and Technology of China, Hefei 230026, China
[2]School of the Gifted Young, University of Science and Technology of China, Hefei 230026, China
[3]Department of Electrical and Electronic Engineering, University of Hong Kong, Pokfulam Road, Hong Kong, China
[4]Department of Optics and Optical Engineering, University of Science and Technology of China, Hefei 230026, China
[5]Department of Thermal Science and Energy Engineering, University of Science and Technology of China, Hefei 230026, China
[6]Physics Experiment Teaching Center, University of Science and Technology of China, Hefei 230026, China.
[7]email: lrd@ustc.edu.cn.
[8]email: huangk17@ustc.edu.cn.

**Perfect optical vortex (POV) is a type of vortex beam with an infinite thin ring and a fixed radius independent of its topological charge. Here we propose the concept of generalized perfect optical vortex along arbitrary curves beyond the regular shapes of circle and ellipse. Generalized perfect optical vortices also share the similar properties as POVs, such as defined only along infinite thin curves and owning topological charges independent of scales. Notably, they naturally degenerate to the POVs and elliptic POVs along circles and ellipses, respectively. We also experimentally generated the generalized perfect optical vortices through a digital micromirror device (DMD) and measured the phase distributions by interferometry, exhibiting good agreements with the simulations. Moreover, we derive a proper modified formula to yield the generalized perfect optical vortices with uniform intensity distribution along predesigned curves. The generalized perfect optical vortices might find the potential applications in optical tweezers and communication.**

Optical vortex beams with the helical wavefronts carry the orbital angular momentum (OAM)[1] of $l\hbar$ per photon ($l$ is the topological charge and $\hbar$ is the reduced Planck constant), which was recognized by Allen et al. in 1992[2]. The pioneering work has excited the intense researches about OVs in various applications including manipulation of micro-particles[3], optical communication[4, 5], quantum information[6-8], plasma diagnostics[9-11], optical imaging and probing[12], and plasmonics[13, 14]. Among these applications, the widely used vortex beams, such as Laguerre-Gaussian and Bessel beams, have the radii of their annular rings determined by their topological charges. As a topological invariant, the topological charge is the most important factor protecting the radial structure of an OV beam from the environmental disturbances, which is known as self-healing effect[15, 16]. Such a topologically dependent intensity profile exists in all the optical vortex beams and thus leads to the inconvenience when it works in a limited field of view, e.g., the tight focusing case by an objective lens with a high numerical aperture.

To overcome this disadvantage, a new type of OVs, termed as perfect optical vortex (POV) beams, was proposed in 2013[17]. The POV beams have their intensity profiles with Dirac delta function independent of topological charges. They can be described mathematically as $\delta(r - r_0) \, exp(il\theta)$, where $(r, \theta)$ is the polar coordinate and $l$ represents the topological charge of the POV. Thus, one can separately control the phase and intensity structures of the OVs. By using diffractive optical elements[18], spatial light modulator (SLM)[17], digital micro-mirror device (DMD)[19], some fruitful progresses such as vector[20] and quantum[6] POVs have been achieved, facilitating the applications in optical tweezers[21], non-diverging speckles[22] and optical free-space communication[4]. Very recently, the circular shape of a POV beam has been extended to an elliptical one by employing the Fraunhofer diffraction of an elliptic Bessel beam[23]. This approach offers a new sight to generalize the POV beam with a non-circular or arbitrary shape with the help of optical Fourier transform. The arbitrarily curved POV beam is important in optical manipulation because it enables the dynamic control of micro-particles by using a single-shot beam, thus avoiding the scanning mode in traditional trapping strategies. There are two fundamental challenges to obtain such generalized POV beams. Firstly, generalized POVs should be predesigned with infinitesimal width along arbitrary smooth trajectories. Secondly, the topological charges and local phase gradient of generalized POVs can be freely controlled without any dependence on the curves.

In this letter, we propose theoretically and demonstrate experimentally the generalized perfect optical vortices (GPOVs) along arbitrary curves beyond the circle and ellipse, solving the problem of extending the concept of perfect optical vortex. To realize the GPOVs experimentally, the required amplitude and phase profiles are simultaneously encoded on DMD as binary holograms, showcasing the GPOVs with circular, elliptic, astroid, Archimedean spiral and "elephant" trajectories. Moreover, we proposed a modified formula of GPOVs to obtain a uniform distributed trajectory in both theory and experiments.

We define GPOVs as a type of vortex beams whose intensity profiles obey Dirac delta function along arbitrary curves and are independent of their topological charges. We firstly consider a smooth 2D curve in Cartesian coordinates: $\vec{c}_2(t) =$

$(x_0(t), y_0(t))$, $t \in [0, T]$, which can be further simplified by employing a specially designed spatial transform:

$$\begin{cases} x = px_0(q) \\ y = py_0(q) \end{cases} \quad (1)$$

where $(p, q)$ are the indices of a new curvilinear coordinate system. To permit an inverse transformation, we also set the Jacobian determinant $|J(p, q)|$ of the conversion to be non-zero:

$$|J(p, q)| = p[x_0(q)y_0'(q) - y_0(q)x_0'(q)] \neq 0 \quad (2)$$

Thus the predesigned curve $\vec{c}_2(t)$ can be written as:

$$\begin{cases} p = 1 \\ q = t \end{cases} \quad (3)$$

In this new curvilinear coordinate system, we could finally come up with the exact mathematical formula of GPOVs according to the definition of GPOVs:

$$C(x, y, z = 0 | \vec{c}_2(t), t \in [0, T]) = \delta(p(x,y) - 1) \, exp\left[ i\sigma \int_0^{q(x,y)} |J(1, \tau)| d\tau \right] \quad (4)$$

where $(x, y, z)$ is the cartesian coordinate, $\delta$ is Dirac delta function, $(p(x,y), q(x,y))$ is the position vector on the curvilinear coordinate plane defined in Eq. (1), $\sigma$ is the key parameters controlling the phase gradient and topological charge of GPOVs along the curve, and $C$ is the complex amplitude of GPOVs in spatial domain. The delta function represents that GPOVs own infinitely thin profiles along given curves. The phase term is represented in terms of a curve invariant proportional to the oriented area of the sector swept in tracing the curve, which means the beam is defined by the curve as a geometric object on the plane and, in particular, is independent of its parameterization.

The optical angular moment (OAM) of GPOVs is related to their topological charge, which is extremely important both for theory and applications. We also calculated the total topological charge $l$ of the GPOVs:

$$l(x, y | \vec{c}_2(t), t \in [0, T]) = \frac{\sigma}{2\pi} \int_0^T |J(1, \tau)| d\tau \quad (5)$$

The integrated term in Eq. (5) is also a curve invariant defined by the curve as a geometric object. The local phase gradient and topological charge of GPOVs can be freely controlled by the parameter $\sigma$, which is independent of their intensity scales.

Next we exemplify a type of GPOVs along elliptic curves: $\vec{c}_2(t) = (a\cos t, b\sin t), t \in [0, 2\pi]$, where $a$, $b$ are the semi-major axis and semi-minor axis of ellipse, is discussed in details here. By using the coordinate transform, the GPOVs have the form:

$$C(x, y, z = 0 | Ellipse) = \delta(p - 1) \, exp(i\sigma abq) \quad (6)$$

where $p = \sqrt{\frac{x^2}{a^2} + \frac{y^2}{b^2}}$, $q = arg\,(bx + iay)$. We can rewrite Eq. (6) in a familiar form:

$$C(x, y, z = 0 | Ellipse) = r_0 \delta(r - r_0) \, exp(il\theta) \quad (7)$$

where $r = \sqrt{\frac{b}{a}x^2 + \frac{a}{b}y^2}$, $\theta = arg\,(bx + iay)$, $r_0 = \sqrt{ab}$, $l = \sigma ab$. Eq. (7) is the exact expression of elliptic perfect vortex (EPOV)[24]. Thus we have proved that EPOVs are just GPOVs along elliptic curves. It therefore confirms the validity of the concept of GPOVs.

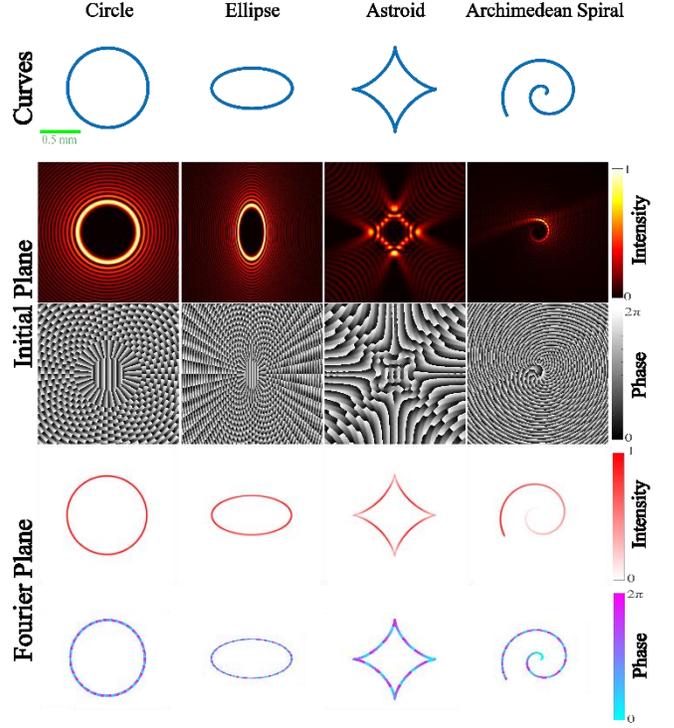

**Fig. 1.** The normalized intensity and phase distribution of GPOVs in frequency domain (initial plane) and GPOVs in spatial domain (Fourier plane) with several curves via simulation.

Due to the infinitely thin intensity profiles along curves and complex expression of GPOVs in spatial domain, it is a common approach to investigate them in frequency domain similar to POVs and EPOVs[23-25]. In a practical setup, the GPOVs in frequency domain are placed at the initial plane. After pass through a Fourier lens with focal length of $f$, the GPOVs in spatial domain are generated on the Fourier plane. Therefore, we can derive the complex amplitude of GPOVs in frequency domain through inverse Fourier transform:

$$Y(x_1, y_1 | \vec{c}_2(t), t \in [0, T]) = \frac{1}{\lambda^2 f^2} \iint_{R^2} C(x, y, z = 0) \, exp\left[ i \frac{2\pi}{\lambda f}(x_1 x + y_1 y) \right] dx dy \quad (8)$$

where $(x_1, y_1)$ is the position vector on the initial plane, and $(x, y)$ is the position vector on the Fourier plane, and $Y$ is the complex amplitude of GPOVs in frequency domain. After substituting the equation of curve, Eq. (8) can be simplified further:

$$Y(x_1, y_1 | \vec{c}_2(t), t \in [0, T]) = \frac{1}{\lambda^2 f^2} \int_0^T \Phi_Y(x_1, y_1, t) |J(1, t)| dt \quad (9)$$

where

$$\Phi_Y(x_1, y_1, t) = exp\left[ i \frac{2\pi}{\lambda f}(x_1 x_0(t) + y_1 y_0(t)) + i\sigma \int_0^t |J(1, \tau)| d\tau \right]$$

Note that this formula is expressed in Cartesian coordinates on the initial plane, which is easy to realize both in simulation and experiments.

In addition to POVs and EPOVs, we also simulate two other types of GPOVs: astroid GPOV and Archimedean GPOV, as shown in Fig. 1. These GPOVs have extremely sharp intensity profiles along arbitrary curves $\vec{c}_2(t)$ with scales independent of topological charges. We can customize the scales and topological charges separately. Besides, the simulated Archimedean GPOV implies that our strategies are valid even for open curves, which could significantly extend the territory of POVs.

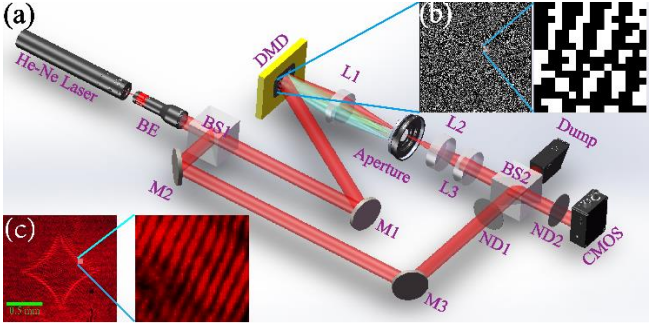

**Fig. 2.** The configuration of our experiments. (a) Optical setup; (b) Binary hologram of astroid GPOV, inset is an enlarged view of the pattern; (c) Interference pattern between an astroid GPOV and a Gaussian beam, inset is an enlarged view of the interferometric fringe.

To demonstrate the proposed GPOVs experimentally, we built up the setup (Fig. 2(a)) by using DMD (DLP Discovery 4100, Texas Instruments, $1024 \times 768$). The He-Ne laser (HRS015B, Thorlabs, 632.8 $nm$) was collimated through a beam expander (BE, $\times 10$) and a mirror $M_1$ to illuminate the DMD. A lens $L_1$ ($f_1$=200 mm) collected the light reflected from the DMD and focused it into a pinhole which selected the 1st-order diffraction. A telescope consisted of lenses $L_2$ and $L_3$ projected the generated GPOVs onto a CMOS camera. Two neutral density filters attenuated the intensity of obtained beams before the CMOS camera. An optical dump is used to collect undesired background light. The generated binary mask for astroid GPOV is shown in Fig. 2 (b) as an example. By projecting the binary holograms onto the DMD, we were able to create steady GPOVs at the CMOS plane ($z = 0$). To encode the complex field, the super-pixel method is used to achieve full control over the spatial phase and amplitude[26, 27]. Such a configuration has sufficiently high resolution for encoding binary holograms. In our experiment, we used a 6-f arrangement to obtain the sufficiently wide curve that was detectable for the CMOS camera. Fig. 3(a) and (b) show the experimental intensity profiles of elliptic and astroid GPOVs, respectively. The good agreement between simulated and measured intensity implies the validity of the proposed GPOVs.

In order to measure the phase of the generated GPOVs in experiments, we applied a method by creating the interference pattern between GPOVs and Gaussian beams as sketched in Fig. 2(a), where a Mach-Zehnder interferometer are established[18, 28]. Two beam splitters (BS1 and BS2) were used to generate the object beam (GPOVs) and reference beam (Gaussian beam), which are tuned into two co-propagating beams with a slight oblique angle for generating the interference patterns, as recorded by the CMOS camera. In Fig. 2(c), we show the recorded interference pattern with clear fringes for the case of astroid GPOV.

Using the interference patterns, we could recover the exact phase profiles of GPOVs. Fig. 3(c) and (d) show the experimentally recovered phase profiles of elliptic GPOV and astroid GPOV, respectively. The phase gradient along the curves within purple dashed lines manifests that the generated GPOVs preserve the designed phase cycles of 30 (ellipse) and 11 (astroid). The curved boundaries between phase cycles is due to a defocus effect in experiments. Both experimental results agree well with the simulations. Furthermore, we also observed the propagation of the GPOVs, as shown in Fig. 3(e). The GPOVs were focused into extremely thin curves at $z = 0$ plane. In this configuration, they could not preserve their intensity profiles at the out-of-focus plane, because the reconstructed beam was designed at the Fraunhofer region.

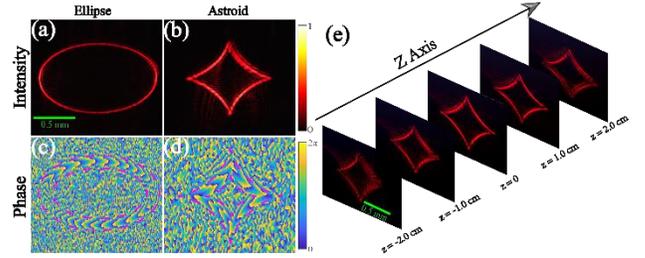

**Fig. 3.** (a), (b), (c) & (d) The experimental intensity and phase distribution of GPOVs with elliptic and astroid trajectories; (e) The intensity distribution of the GPOV with astroid trajectory along z axis.

Note that both the simulated and obtained intensity profiles are not uniform along the trajectories except circles. This inhomogeneous effect becomes severe especially at the cusp points along the curves, see Fig. 1 for example. This phenomenon originates from the unique properties of Dirac delta function[29]. In both numerical simulations and experiments, As shown in Eq. (4), the GPOVs are present in a specially designed spatial transform described in Eq. (1) instead of the Cartesian coordinates. This definition actually modulates the amplitude distribution of GPOVs along the curves. This phenomenon has been reported in EPOVs, where a proper designed elliptical aperture is suggested to realize a uniform amplitude distribution[23]. Unfortunately, it is quite difficult to design such apertures for complex curves.

To further extend the potential in applications of GPOVs, we need to generate GPOVs with uniform intensity along given curves. Here we propose a modification on the expression of GPOVs, which can be treated as applying a predesigned amplitude modulation on GPOVs along the curves by the ratio of the $|\vec{c}_2'(q)|$ term to $|J(1,q)|$ term:

$$C_{Modified}(x, y, z = 0 | \vec{c}_2(t), t \in [0, T]) =$$
$$\frac{|\vec{c}_2'(q)|}{J(1,q)} \delta(p-1) \, exp\left[\frac{i\sigma}{\omega_0^2} \int_0^q J(1,\tau)d\tau\right] \quad (10)$$

where $|\vec{c}_2'(q)| = \sqrt{x_0'^2(q) + y_0'^2(q)}$. Similar to Eq. (8), we derive the predesigned amplitude modulation for GPOVs in frequency domain:

$$Y_{Modified}(x_1, y_1 | \vec{c}_2(t), t \in [0, T]) = \quad (11)$$

$$\frac{1}{\lambda^2 f^2}\int_0^T \Phi_Y(x_1,y_1,t)|\vec{c}_2'(t)|dt$$

The modified formulas Eqs. (10)-(11) not only has exactly the same expression as the EPOVs with predesigned elliptical aperture in elliptic case, but also are suitable for complex curves.

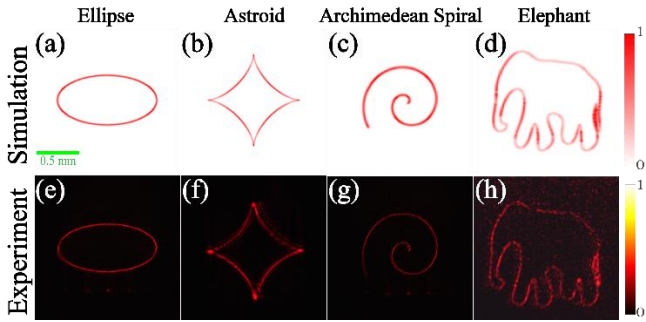

**Fig. 4.** The simulated and experimental intensity distribution using revised formula under finite aperture.

Fig. 4 shows the simulated and experimental intensity distribution of GPOVs under the predesigned amplitude modulation described by Eqs. (10)-(11). All the shapes including elliptical, astroid and Archimedean curves can be realized experimentally, as shown in Figs. 4(e)-(g). In addition, an elephant with four feet and one nose can also be created with good profile in Fig. 4(h), which is highly consistent with the simulation in Fig. 4(d). Compared with Fig. 1, the revised formula yields better uniformity for intensity profiles along the curves. The advantage of this revision is to acquire uniform intensity distribution, although it is difficult to realize completely uniform curves at cusp points such as four vertexes in Fig. 4 (b) and (f). By using this revision, we can overcome the key difficulties in applying the GPOVs in the fields such as optical trapping and single shot lithography[30].

In conclusion, we have proposed the concept of GPOVs along arbitrary curves. GPOVs with varied shapes possess the similar properties with POVs, such as defined only along infinitely thin curves and owning topological charges independent of scales. Note that, they naturally degenerate to the POVs and elliptic POVs in circle and ellipse cases, respectively. For applications such as optical trapping and single shot lithography, we give a proper modified formula of GPOVs to produce uniform intensity along predesigned curves. We also experimentally generated the GPOVs through DMD and measured their phase profiles by interfering with Gaussian beams. The experimental results are consistent with the simulations. These vortex beams are valuable in micromanipulation, quantum communication, optical imaging, and single shot lithography.

**Funding.** This work is sponsored by the National Natural Science Foundation of China (31670866, 60974038, 61875181, 61705085); Natural Science Foundation of Anhui Province (1708085MF143); CAS Pioneer Hundred Talents Program.

**Disclosures.** The authors declare no conflicts of interest.